\documentstyle[emulateapj,apjfonts,epsfig]{article}

\newcommand{\xtegsfc}{XTE~J1751-305}
\newcommand{\xtemit}{XTE~J0929-314}

\begin{document}

\title{Hot White Dwarf Donors in Ultracompact X-Ray Binaries}

\author{Lars~Bildsten}
\affil{\footnotesize Kavli Institute for Theoretical Physics and
Department of Physics\\
  Kohn Hall, University of California, Santa Barbara, CA 93106;
   bildsten@kitp.ucsb.edu}

\begin{abstract}

The discovery of two accreting millisecond X-ray pulsars in binaries
with $\approx 43$ minute orbital periods allows for a new probe of the
donor's structure. For \xtegsfc, only a hot white dwarf (WD) can fill
the Roche Lobe. A cold He WD is a possible solution for \xtemit,
though I will show that evolutionary arguments make a hot WD more
likely. In addition to being larger than the $T=0$ models, these
finite entropy, low-mass ($M_c<0.03M_\odot$) WDs have a minimum mass
for a fixed core temperature. If they remain hot as they lose mass and
expand, they can ``evaporate'' to leave an isolated millisecond radio
pulsar. They also adiabatically expand upon mass loss at a rate faster
than the growth of the Roche radius if the angular momentum deposited
in the disk is not returned to the donor.  If the timescale of the
resulting runaway mass transfer is shorter than the viscous timescale
in the outer disk, then the mass transfer instability of Ruderman and
Shaham for He WDs would be realized. However, my estimates of these
timescales still makes the instability unlikely for adiabatic
responses. I close by noting the possible impact of finite $T$ WDs on
our understanding of AM CVn binaries.

\end{abstract}

\centerline{\bf TO APPEAR IN ASTROPHYSICAL JOURNAL LETTERS}

\keywords{accretion, accretion disks --- binaries: close --- 
stars: low-mass, brown dwarfs  --- stars: neutron --- white dwarfs ---
X-rays:binaries}

\section{INTRODUCTION}

 The recent discovery by the {\it Rossi X-Ray Timing Explorer} of two
accreting millisecond pulsar transients \xtegsfc \ ($\nu_s\approx 435
\ {\rm Hz}$; Markwardt et al. 2002) and \xtemit \ ($\nu_s\approx 185 \
{\rm Hz}$; Galloway et al. 2002) has allowed us to learn about the
donors in these ultracompact binaries. At $P_{\rm orb}\approx 43$
minute orbital periods, H-rich donors are ruled out (Nelson, Rappaport
\& Joss 1986), and He-rich stars (see Podsiadlowski, Rappaport \&
Pfahl 2002 for an updated discussion of the range of H/He ratios) or
WDs are filling the Roche lobe (RL). If a degenerate WD, then the
orbital period increases as mass is transferred at the rate set by
angular momentum losses from gravity waves, $\dot J_{\rm GR}$
(see Verbunt 1993).

 The measured pulsar orbital parameters yield the RL filling companion's
mass, $M_{\rm c}$, and the minimum mass transfer rate, $\dot M_{\rm
GR}=3M_{\rm c}\dot J_{\rm GR}/2J$, which are shown in Figure~1 for
neutron stars of $M_{x}=1.4-2.0 M_\odot$.  The low $\dot M$'s 
are the likely cause for the transient behavior, as the
steady-state outer disk temperature 
is below these element's ionization temperature (Tsugawa \& Osaki
1997; Menou et al. 2002), even if X-ray heating is included at the rate
inferred in other X-ray binaries (Dubus, Hameury \& Lasota 2001). If
identical WDs are the donors in both of these binaries, then their
identical orbital periods would require the same RL filling
solutions. This would constrain the inclination for \xtemit \ to
less than 37 degrees and $M_c>0.013M_\odot$ for both \xtemit \ and
\xtegsfc.

 The dotted lines in Figure 2 show the mass-radius relation for RL
filling donors of \xtegsfc\ (lower line) and \xtemit\ (upper
line). The solid and dashed lines (which exhibit a maximum radius due
to the onset of Coulomb physics) are the cold ($T=0$) WDs of Zapolsky
\& Salpeter (1969, hereafter ZS) for pure He and C, respectively. A
cold He WD will fill the RL for \xtemit \ (Galloway et al. 2002),
whereas there are no cold WD solutions that fill the RL for \xtegsfc \
(Markwardt et al. 2002).

 This evidence for finite $T_c$ WDs motivated my calculations
shown in Figure 2 by the solid and dashed lines
that diverge at low $M_c$. These models (see \S 2) are for He
at $T_c=10^5\ {\rm K}$ and $10^6 \ {\rm K}$ and C at
$T_c=10^6\ {\rm K}$ and $T_c=3\times 10^6 \ {\rm K}$. These provide RL
filling solutions, but don't {\it a priori} differentiate between He
or C or any other element that might dominate the donor
star. I consider C WDs throughout this paper, as Schulz et
al. (2001) have measured a high Ne to O ratio in the matter
transferred onto NSs in ultracompact X-ray binaries (confirmed
by Juett, Psaltis \& Chakrabarty 2002; Homer et al. 2002; Juett \&
Chakrabarty 2002). These measurements led Schulz et al. (2001) and
Juett et al. (2001) to suggest that the donors in these binaries 
are the cores of previously crystallized C/O WDs.

In \S 2 I justify (on evolutionary grounds) that hot WDs are expected,
describe the construction of their $R_c(M_c)$
relations, and point out the existence of a minimum $M_c$ solution for a
given $T_c$.  I put the models in the context of mass
donors in \S 3 and calculate
\begin{equation}
\label{eq:nad}
n_{\rm Ad}={d\ln R_c\over d\ln M_c},
\end{equation} 
the adiabatic exponent needed to evaluate the stability of mass
transfer. Though uncertainties remain regarding the ability of the
accretion disk to store angular momentum on long timescales (Verbunt
\& Rappaport 1988), my initial work finds that the expansion of a He
WD donor under mass transfer can exceed that of the RL, possibly
allowing for the Ruderman \& Shaham (1983) instability.

\section{FINITE ENTROPY WHITE DWARFS AND EVAPORATION}
 
The material that fills the RL in these systems today was once the
deep interior of an initially more massive WD, so I begin by
discussing the extent to which prior evolution (before RL filling) can
affect the temperature of the current low-mass WDs.  There are a few
evolutionary scenarios that place a He or C/O WD in a tight enough
orbit (following a common-envelope induced spiral-in) about a NS that
gravity wave emission will place it in contact within 5-10 Gyr (Rasio,
Pfahl \& Rappaport 2000; Yungelson, Nelemans \& van den Heuvel
2002). In this case the WD has had some time to cool before initiating
RL overflow. Nelemans et al. (2001) discuss a similar scenario for
the origin of AM CVn binaries.

 The rate at which they cool differentiates He and C/O WDs.  The
larger specific heat of a He WD slows  its cooling (Althaus \&
Benvenuto 1997; Hansen \& Phinney 1998). For example, a $0.20 M_\odot$
He WD would have $T_c=10(3.3)\times 10^6 \ {\rm K} $
at $\approx 1.0(4.0)$ Gyr (Althaus \& Benvenuto 1997) and be 
liquid when RL filling occurs. The smaller
specific heat of a $0.6M_\odot$ C/O WD allows it to cool
to $2.5\times 10^6 {\rm K}$ in 4 Gyr and begin crystallization
(Salaris et al. 2000). Hence, most C/O WDs would be solid prior to RL
filling (this is needed for the enhancement of $^{22}{\rm Ne}$ by
fractionation; Schulz et al. 2001) and might only melt if heated
during the GW inspiral (Iben, Tutukov \& Fedorova 1998).  The other
differentiation between such WDs comes from their mass, as Yungelson
et al. (2002) argue that the mass transfer at the onset of RL
filling will be unstable if $M_c>0.44M_\odot$, 
excluding traditional C/O WDs from ever reaching short orbital
periods. However, Yungelson et al. (2002)
note that ``hybrid'' WDs with C/O cores can have $M_c<0.44
M_\odot$ and thus evolve to lower mass systems filling the RL at 20-40
minutes.

 The inability for the donor WDs to cool
on the $\sim$ Gyr timescale of the mass transfer phase (Rappaport et
al. 1987) makes it clear that the initial entropy is the minimum
value attainable. All of the entropy is in the liquid ions, 
which have an adiabatic scaling, $T_c\propto \rho_c^{0.5-0.6}$
(Hernanz et al. 1988), less steep than for an ideal gas ($T\propto
\rho^{2/3}$). Since $\rho_c\propto M_c^2$, a $\approx 0.013M_\odot$ He
WD made by {\it adiabatically expanding} the deep interior of a $0.20
M_\odot$ WD would be a factor of $\approx 15$ cooler, or $T_c\approx
6.6(2.2)\times 10^5 {\rm K}$ for the initial range of $T$'s discussed
earlier. Of course, any tidal heating will increase $T_c$ (see \S 3).

So I have constructed $M_c<0.03M_\odot$ WDs of finite $T_c$
that contains the Coulomb
physics (from Chabrier \& Potekhin 1988) 
that causes the turnover to constant density ``rocks'' in the
mass-radius relation at $\approx 10^{-3} M_\odot$ seen in the ZS models in
Figure 2. I have imposed an arbitrary (but
convectively stable) temperature profile of $T=T_c(P/P_c)^{1/5}$ while
integrating hydrostatic balance and mass conservation. I used 
the electron
equation of state of Paczynski (1983) and halt 
integrations at the point where the pressure has fallen to
$10^{-5}$ of the central value, which avoids the need for an 
envelope model. The results are shown in Figure 2 and make clear
that the temperatures expected from {\it adiabatic evolution of a He
WD} are adequate to provide the slight radius expansion needed to fill
the RL for these accreting millisecond pulsars. This is easier for He
WDs because of the extra number of ions per electron compared to C/O
WDs. White dwarfs made of C/O will need some tidal heating to 
fill these RLs.

 I also found that there is a minimum mass solution for a WD of fixed
$T_c$. This is related to the electron Fermi energy of the nearly
constant density solutions at low masses and is reflected in the
radius divergence of the models in Figure 2 at low $M_c$. My 
semi-analytic modeling finds that these finite $T_c$ solutions
eventually ``turnover'' to follow a track of $R_c\propto M_c$, as
expected for an ideal gas polytrope (where $T_c\propto
M_c/R_c$). Since there is no way to reach these solutions of higher
$M_c$ while mass is being lost, these ``evaporative''
endpoints where $d\ln R_c/d\ln M_c$ diverges can only be reached if
tidal heating is adequate to keep these models hot while the central
density is dropping. More work remains to actually show that this 
will eliminate the companion and lead to an isolated millisecond radio
pulsar. 

\section{MASS TRANSFER INSTABILITIES AND QUIESCENT EMISSION}

 Barring tidal heating adequate to reach the ``evaporative'' endpoint
alluded to in \S 2, I will now discuss a possible 
mass transfer instability.  Matter leaving the WD 
takes angular momentum with it as it settles into the
accretion disk. This angular momentum is returned to the WD
via tidal torques once there has been enough time for viscosity to
move the material outwards from the splash point.  The timescale for
this angular momentum loop to be closed has been considered to be long
enough (or uncertain enough) that Ruderman \& Shaham (1983) raised the
possibility that a mass transfer instability could occur when the
expansion of a low-mass donor ($M_c\ll M_x$) due to mass loss
(measured by $n_{\rm Ad}$ of eq. [\ref{eq:nad}]) exceeds that of the
RL, measured by
\begin{equation}
\label{eq:nr}
n_{\rm R}={d\ln R_R\over d\ln M_c}={d\ln a\over d\ln M_c}+{1\over 3},
\end{equation} 
where $a$ is the orbital separation. 
The value of $n_{\rm R}$ presuming the angular momentum ``sink'' of
the disk (using the fitting formula of Verbunt \& Rappaport 1988) is
shown in Figure 3 by the dotted lines for (from left to right)
$M_x=0.6,1.0,1.4,$ \& $1.8 \ M_\odot$. 

Ruderman \& Shaham (1983) presumed $n_{\rm Ad}=-1/3$, which was shown
to be an overestimate by Hut \& Paczynski (1984) and Bonsema \& van
den Heuvel (1985), who found $n$ from the ZS relation (shown in Figure
3 by the solid[dashed] lines for He[C]). These authors noted that a ZS
model moved the crossing ($n_{\rm Ad}<n_{\rm R}$) to such low $M_c$
that the binary might not reach it in a Hubble time. However, for
finite $T$ He WDs, I show that this conclusion is altered.  The solid
circles are my evaluation of $n_{\rm AD}$ for the $T_c=10^5 \ {\rm K}$
(set closest to the solid line) and $T_c=10^6 \ {\rm K}$ (bottom-most
set) He WDs of Figure 2 and show that the finite
entropy model sits between the $T=0$ value and the ``naive'' $n=-1/3$
guess for a perfect gas.  For a $M_x=1.4 \ M_\odot$ accretor, the
instability question is first raised (since $n_{\rm Ad}<n_{\rm R}$)
when a $10^6 \ {\rm K}$ He WD has $M_c\approx 0.01\ M_\odot$. This
will occur about a Gyr after mass transfer has started presuming just
gravity wave emission. My evaluations of $n_{\rm Ad}$ for C WDs at the
$T_c$'s of Figure 2 never found such a crossing.

The instability question is thus 
raised for He WDs and requires some extra entropy from tidal heating. The
amount of heating is small, as for liquid He to reach 
$T_c\approx 10^6 \ {\rm K}$ from about 1/2 that value (which it had initially)
requires $3k_B(0.5T_c)/4m_p\approx 3\times 10^{13} \ {\rm
erg \ g^{-1}}$ (the specific heat in the liquid state is $\approx
3k_BT$ per ion). This energy is comparable to the current rotational
energy per gram for a tidally locked WD at the 40 minute orbital
period.  Hence, only a fraction of the higher rotational energy per
gram from the tidally locked WD at a shorter orbital period needs to
be deposited to yield $T_c\approx 10^6 \ {\rm K}$ today. 

 The remaining criticism against the mass transfer instability is one
of relative timescales (Verbunt \& Rappaport 1988; Priedhorsky \&
Verbunt 1988). The mass-transfer instability most likely grows on a
timescale $\tau_g\approx M_cH/\dot M_c R_c$ (Verbunt \& Rappaport
1988), where $H$ is the scale height in the WD atmosphere, fixed by
the X-ray heating in quiescence from the hot NS (Bildsten \&
Chakrabarty 2001). The binary spends most of the time undergoing mass
transfer to the outer disk, but little to no accretion onto the
NS. Bildsten \& Chakrabarty (2001) showed that the X-ray emission
always detected from NS binaries in quiescence (see Bildsten \&
Rutledge 2001 for an overview) heat the companion on the side facing
the NS. Presuming the NS thermal emission is at the level predicted by
Brown, Bildsten \& Rutledge (1998) (giving $L_{X,q}\approx 5\times
10^{32} \ {\rm erg \ s^{-1}}$) then the WDs in \xtegsfc \ and \xtemit
\ will have $T_{\rm Eff}\approx 5000-6500 \ {\rm K}$.\footnote{These
high $T_{\rm Eff}$'s should aid detection of the quiescent
counterpart, as the luminosity (presuming $R_c=0.045R_\odot$) will be
$L_c\approx (0.5-1.0)\times 10^{-3} L_\odot$ just from reprocessing
the NS's thermal emission.  Presuming $T_{\rm Eff}\approx 6000 \ {\rm
K}$ and using the WD radius implied by RL filling, I get $M_V\approx
12$ for the heated WD face. For the high latitude source \xtemit \ at
$d\approx 5-7$ kpc (Galloway et al. 2002), this gives $V\approx 26
$. } The scale height is $H/R_c\approx 3\times 10^{-4}$, yielding
$\tau_g\approx 10^6 \ {\rm yr}$. This is still much longer than the
time for viscosity to finally play a role in moving material outwards
in the outer accretion disk, most likely allowing for the angular
momentum to get back to the donor and move it out, so that $n_{\rm
R}=-5/3$. The adiabatic index is never less than $-5/3$. However, the
isothermal WD response is $<-5/3$ and mass transfer would be unstable
for $M_c<4(9.5)\times 10^{-3}M_\odot$ in a He WD of $T_c=4(10)\times
10^5 \ {\rm K}$ as long as internal heating can keep the WD
isothermal as mass is lost. These masses are near the values I noted
earlier for the ``evaporation'', namely where the mass-radius relation
turns over.

\section{CONCLUSIONS}

The timing of two accreting millisecond pulsars (Markwardt et
al. 2002; Galloway et al. 2002) in ultracompact binaries has probed
the WD donor properties to new levels and shown that they are of
finite entropy. This motivated my calculations of low-mass WDs of
finite $T_c$ that allow for $T_c$ to be constrained.  For He WDs, the
implied $T_c$ are nearly that expected just from adiabatic expansion
of the initially hot WD that filled the RL. Only a small amount of
tidal heating is needed. More tidal heating is needed to make a C/O WD
fill the RL.

 These finite $T$ solutions allowed for a re-evaluation of Ruderman \&
Shaham's (1983) scenario for making isolated millisecond radio pulsars
via a mass transfer instability. I find that the adiabatic mass
transfer instability can occur for a hot He WD as long as the angular
momentum leaving the RL filling star is not returned to the orbit. I
have thus eliminated one criticism of their model, though the question
of angular momentum elimination remains a serious one. The physics of
hot, low-mass WDs also yields a minimum mass WD solution for a fixed
$T_c$ so that a mass transfer instability can occur if the donor
remains isothermal under mass loss.  I thus speculate that an
evaporative or mass transfer instability endpoint might occur as long
as tidal (or other) heating persists at 40-80 minute orbital periods.

 This work also impacts AM CVn binaries, where a low mass He star
donates material to a more massive WD (see Solheim 1995 for a
review). The larger WD radii lead to more GW emission and a higher
$\dot M_c$ than expected for a given $P_{\rm orb}$. Hence, models
which track the WD entropy will fall between the degenerate and
non-degenerate models in Nelemans et al. (2001) and depend on both the
age of the system when RL filling occurs (as this fixes the initial WD
entropy) and any tidal heating that occurs during the mass
transfer. If either of the instabilities discussed above occur, then
the endpoint of AM CVn's could well be a DB WD (e.g. Tutukov \&
Yungelson 1996).

\acknowledgments

I thank Deepto Chakrabarty for alerting me to the discovery of these
transients and for many conversations. Ira
Wasserman provided great physics insights during the progress of this
work, which was supported by NASA via grant NAG 5-8658 and by the NSF
under grants PHY99-07949 and AST01-9642. L. B. is a Cottrell Scholar
of the Research Corporation.

\vfill\eject

\begin{figure}
\plotone{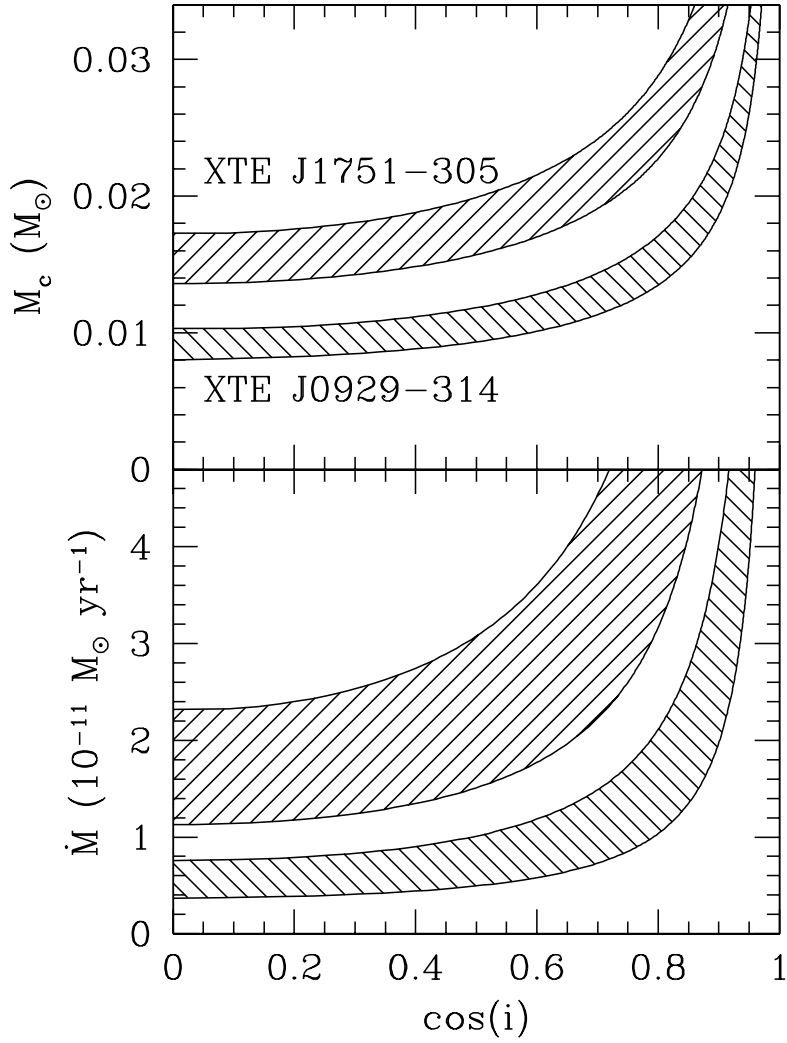}
\figcaption{The companion masses and gravity-wave driven mass transfer rates 
for \xtegsfc\ and \xtemit\ as a function of cosine of the 
inclination angle. The 
hatched regions (upper is \xtegsfc \  and lower is \xtemit) 
are bounded below by $M_x=1.4M_\odot$ and above by
$M_x=2.0M_\odot$. It was presumed that the donor
responds to mass transfer with $n=-1/3$, only a small difference from 
$n_{\rm Ad}$. 
\label{fig:trans}}
\end{figure}

\begin{figure}
\plotone{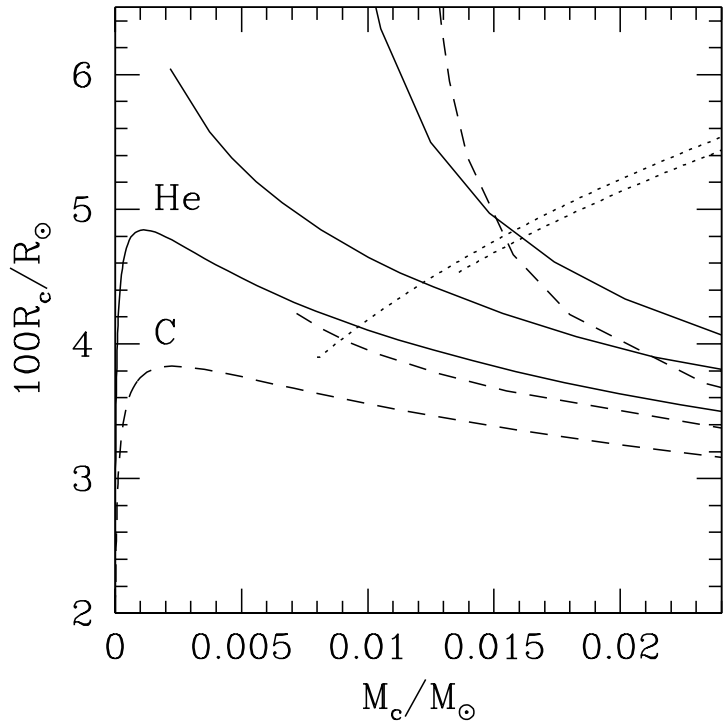} \figcaption{ Roche lobe filling mass-radius
relations for \xtegsfc\ and \xtemit\ compared to WDs. The dotted
lines denote the RL filling donors of \xtegsfc \ (lower line) and
\xtemit \ (upper line). The solid (dashed) lines are $T=0$ WDs of ZS
for pure He (C) using Rappaport \& Joss's (1984) correction to
ZS. The 
lines which diverge at small $M_c$ are hot WDs. The He WDs
have $T_c=10^5\ {\rm K}$ and $10^6 \ {\rm K}$ and the C WDs 
have $T_c=10^6\ {\rm K}$ and $T_c=3\times 10^6 \ {\rm K}$. 
\label{fig:massr}}
\end{figure}

\begin{figure}
\plotone{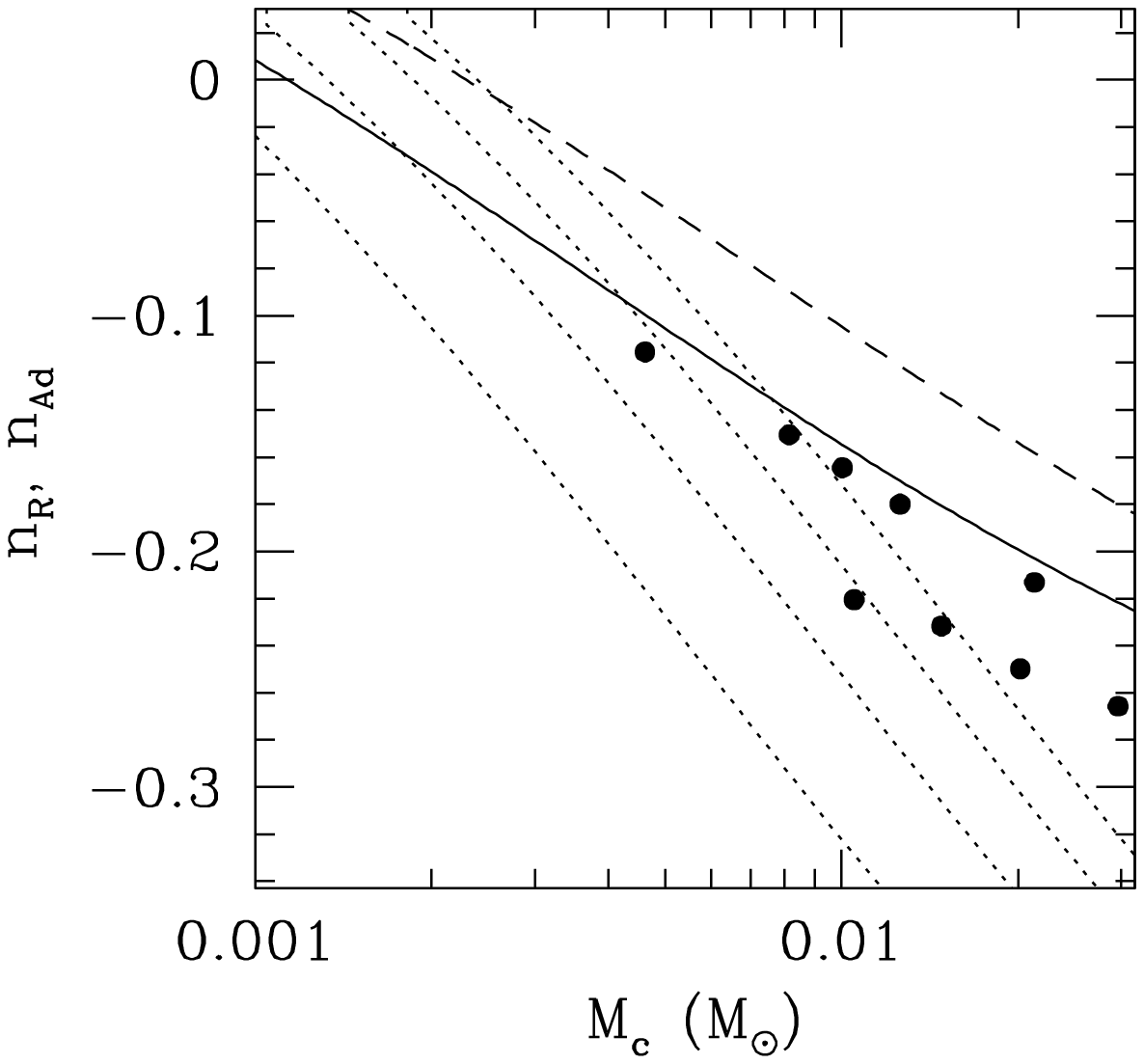}
\figcaption{Roche radius and stellar radius changes with mass
transfer. The dotted lines show $n_{R}$ for (from left to right) 
$M_x=0.6,1.0,1.4\ \& \ 1.8 M_\odot$ when the
transferred angular momentum is stored in the disk. The solid (dashed)
line are $n_{\rm Ad}=n_{\rm ZS}$ for $T=0$ pure He (C) WDs. 
The filled circles are $n_{\rm Ad}$ for He WDs of $T_c=10^5 \ {\rm
K}$ (set closest to the solid line) and $T_c=10^6 \ {\rm K}$. 
\label{fig:unstabs}}
\end{figure} 


\begin{references}

\noindent
Althaus, L. G., \& Benvenuto, O. G. 1997, \apj, 477, 313 

\noindent
Bildsten, L. \& Chakrabarty, D. 2001, \apj, 557, 292 

\noindent
Bildsten, L. \& Rutledge, R. E. 2001, in ``The Neutron-Star Black Hole
Connection'', eds. C. Kouveliotou, J. Ventura \& E. P. J. van den
Heuvel (Dordrecht: Kluwer) p. 245 

\noindent
Bonsema, P. F. J. \& van den Heuvel, E. P. J. 1985, A\&A, 146, L3 

\noindent 
Brown, E., Bildsten, L. \& Rutledge, R.~E. 1998, ApJ, 504, L95

\noindent
Chabrier, G. \& Potekhin, A. V. 1998, Phys. Rev., E58, 4941

\noindent
Dubus, G., Hameury, J.-M., \& Lasota, J.-P. 2001, A\&A, 373, 251. 

\noindent
Galloway, D. K., Chakrabarty, D., Morgan, E. H.,
\&  Remillard, R. A., 2002, to appear in Ap J Letters 

\noindent
Hansen, B. M. S. \& Phinney, E. S. 1998, MNRAS, 294, 557 

\noindent
Hernanz, M., Isern, J., Canal, R., Labay, J. \& Mochkovitch, R. 1988,
\apj, 324, 331 

\noindent
Homer, L., Anderson, S. F., Wachter, S. \& Margon, B., 2002, submitted
to \apj, astro-ph/0205332 

\noindent
Hut, P. \& Paczynski, B. 1984, \apj, 284, 675 

\noindent
Iben, I., Tutukov, A. V. \& Fedorova, A. V. 1998, \apj, 503, 344 

\noindent
Juett, A. M., Psaltis, D. \& Chakrabarty, D. 2001, \apj, 560, L59 

\noindent
Juett, A. M. \& Chakrabarty, D. 2002, submitted to \apj,
astro-ph/0206417 

\noindent 
Markwardt, C. B., Swank, J. H., Strohmayer, T. E., in 't Zand,
J. J. M. \& Marshall, F. E. 2002, \apj, 575, L21

\noindent
Menou, K., Perna, R. \& Hernquist, L. 2002, \apj, 564, L81 

\noindent
Nelemans, G., Portegies Zwart, S. F., Verbunt, F. \& Yungelson,
L. R. 2001, A\& A, 368, 939 

\noindent
Nelson, L. A., Rappaport, S. A. \& Joss, P. C. 1986, \apj, 304, 231 

\noindent 
Paczynski, B. 1983, \apj, 267, 315 

\noindent
Podsiadlowski, P., Rappaport, S. \& Pfahl, E. D. 2002, \apj, 565, 1107

\noindent
Priedhorsky, W. C. \& Verbunt, F. 1988, \apj, 333, 895 

\noindent
Rappaport, S. \& Joss, P. C. 1984, \apj, 283, 232 

\noindent
Rappaport, S., Nelson, L. A., Ma, C. P. \& Joss, P. C. 1987, \apj,
322, 842 

\noindent 
Rasio, F. A., Pfahl, E. D. \& Rappaport, S. 2000, \apj, 532, L47 

\noindent
Ruderman, M. \& Shaham, J. 1983, Nature, 304, 425 

\noindent
Salaris, M., Garcia-Berro, E., Hernanz, M., Isern, J. \& Saumon,
D. 2000, \apj, 544, 1036 

\noindent
Schulz, N. S., Chakrabarty, D., Marshall, H. L., Canizares, C. R.,
Lee, J. C. \& Houck, J. 2001, \apj, 563, 941 

\noindent
Solheim, J.-E. 1995, Baltic Astron., 4, 363 

\noindent
Tsugawa, M. \& Osaki, Y. 1997, PASJ, 49, 75 

\noindent
Tutukov, A. \& Yungelson, L. 1996, MNRAS, 280, 1035 

\noindent
Verbunt, F. \& Rappaport, S. 1988, \apj, 332, 193 

\noindent
Verbunt, F. 1993, ARAA, 31, 93 

\noindent
Yungelson, L. R., Nelemans, G. \& van den Heuvel, E. P. J., 2002,
A\&A, 388, 546 

\noindent
Zapolsky, H. S. \& Salpeter, E. E. 1969, \apj, 158, 809 (ZS) 

\end{references}
\end{document}